\documentclass[aps,pra,twocolumn,groupedaddress,showpacs,showkeys]{revtex4}

\usepackage{amsfonts}
\usepackage{amsmath}
\usepackage{amssymb}

\allowdisplaybreaks

\begin{document}


\title{Generalized model-independent approach to intrinsic decoherence}


\author{D. Salgado}
\email[]{david.salgado@uam.es}
\affiliation{Dpto. F\'{\i}sica Te\'{o}rica, Universidad Aut\'{o}noma de Madrid\\
28049 Cantoblanco, Madrid (Spain)}

\author{J.L. S\'{a}nchez-G\'{o}mez}
\email[]{jl.sanchezgomez@uam.es}
\altaffiliation{Permanent Address}
\affiliation{Dpto. F\'{\i}sica Te\'{o}rica, Universidad Aut\'{o}noma de Madrid\\
28049 Cantoblanco, Madrid (Spain)}


\date{\today}

\begin{abstract}
A formalism is presented to express decoherence both in the markovian and nonmarkovian regimes and both dissipative and nondissipative in isolated systems. The main physical hypothesis, already contained in the literature \cite{Mil91a,BonOliTomVit00a}, amounts to allowing some internal parameters of the system to evolve in a random fashion. This formalism may also be applicable to open quantum systems. 
\end{abstract}

\pacs{03.65.Yz,02.50.-r}
\keywords{Decoherence, Markovianity, Dissipation, Randomness}

\maketitle



Decoherence is at the heart of the destruction of quantum superpositions in most physical systems. Its role is basic in questions such as the quantum-classical transition \cite{Zur91a}, the realization of quantum computation or the quantum processing of  information \cite{ChuNie00a}. From a quantitative point of view the effect of decoherence is commonly described by the action of the environment upon the physical system of interest \cite{GiuJooKieKupStaZeh96a,Zur82a}. This view inevitably depends on the chosen modelization of this environment. Recently \cite{BonOliTomVit00a} (see also \cite{Bon99a}) a model-independent approach to decoherence has been initiated based on the hypothesis of randomness of some of the internal parameters of the system, thus reducing the role of the environment. Here we present a triple generalization of this model-independent approach which  allows us to find more general master equations both in the markovian and in the nonmarkovian regimes\footnote{As usual, markovianity is understood in the sense of an evolution which satisfies a semigroup condition.} and both dissipative and nondissipative.

Randomness in an isolated system is introduced by claiming that the evolution time is a stochastic parameter \cite{BonOliTomVit00a,Bon99a} (previous references in this respect are \cite{Mil91a,MoyBuzKimKni93a}). This physical hypothesis is implemented by noticing that the experimental observations are described by the time-averaged density operator, as expressed by the relation 

\begin{equation}
\label{BonDensOp}
\bar{\rho}(t)=\int_{0}^{\infty}dt'P(t,t')\rho(t')
\end{equation}

\noindent where $P(t,t')$ denotes the probability density associated with the random time $t'$ and $\rho(t')$ denotes the usual quantum density operator. To find an expression for $P(t,t')$ two conditions are imposed \cite{BonOliTomVit00a}: i) $\bar{\rho}(t)$ must be a density operator itself; ii) the evolution of $\bar{\rho}(t)$ must be markovian. Under these assumptions and using the integral definition of the Gamma function, one arrives at a fixed probability density $P(t,t')$ (up to two parameters) and thus at a fixed time-averaged density operator $\bar{\rho}(t)$. Here we drop condition ii), thus comprising nonmarkovian evolutions and moreover using different tools we also allow for distinct markovian situations. The nonmarkovian regime is interesting in its own since it is the rigorous type of dynamics (prior to approximations) followed by a physical system coupled to an environment \cite{AlickiLendi87a,GorFriVerKosSud78a}. 
Furthermore the experimental advances in mesoscopic scales (for instance, in creating atom lasers with Bose-Einstein condensates \cite{MoyHopSav98a}) also demands a better understanding of this regime. The generality obtained also allows us to embrace dissipative decoherence in contrast to \cite{BonOliTomVit00a,Mil91a}.

The idea is to express the evolution operator $U(t)$ using the spectral decomposition theorem for unitary operators \cite{Kreyszig78a} and then introduce the randomness in the time evolution. Let us recall then that 

\begin{equation}\label{UnitOp}
U(t)=\int_{-\pi}^{\pi}e^{-i\theta t}dE_{\theta}
\end{equation}

\noindent where $E_{\theta}$ is the spectral family on $[-\pi,\pi]$ associated to $U(t)$. The randomness hypothesis can be introduced in different ways. The most na\"{\i}ve way is just to substitute $t$ in expression (\ref{UnitOp}) for a random variable $\chi_{t}$ which expresses the stochastic nature of the evolution. The properties of $\chi_{t}$ (distribution, moments, \dots) will depend on the characteristics of the physical system under study. To grasp this generality it is convenient to consider  $\chi_{t}$ as the solution to an Ito stochastic differential equation \cite{Oksendal98a}:

\begin{equation}
d\chi_{t}=b(t,\chi_{t})dt+\sigma(t,\chi_{t})d\mathcal{B}_{t}
\end{equation} 

\noindent where $b(\cdot,\cdot)$ and $\sigma(\cdot,\cdot)$ are real functions of two real variables and $\mathcal{B}_{t}$ denotes standard real brownian motion. These two functions encode the information of the specific physical system. As a first example we may consider $b(t,\chi_{t})=1$ and $\sigma(t,\chi_{t})=\sigma(t)$ (with clear abuse of notation). This allows us to write $\chi_{t}=t+\int_{0}^{t}\sigma(s)d\mathcal{B}_{s}$, hence the expectation value of $\chi_{t}$ is the standard deterministic time $t$ and its deviation from it is given by $\lambda(t)\equiv\int_{0}^{t}\sigma^{2}(s)ds$. This can be considered a weak-coupling-like limit. With these hypotheses, we may construct a random evolution operator given by

\begin{equation}\label{RandEvolOp}
U_{\text{rand}}(t)=\int_{-\pi}^{\pi}e^{-i\theta t-i\theta\int_{0}^{t}\sigma(s)d\mathcal{B}_{s}}dE_{\theta}
\end{equation}

Following \cite{BonOliTomVit00a} the density operator describing experimental results will be given by

\begin{equation}\label{ExpVal}
\bar{\rho}(t)=\mathbb{E}[\rho(t)]=\mathbb{E}[U_{\text{rand}}(t)\rho(0)U_{\text{rand}}^{\dagger}(t)]
\end{equation}

\noindent where $\mathbb{E}$ denotes expectation value with respect to the probability distribution of $\chi_{t}$. Using Ito calculus \cite{Oksendal98a}, the expectation value $\mathbb{E}[e^{-i(\theta-\theta')\int_{0}^{t}\sigma_{s}d\mathcal{B}_{s}}]=e^{-\frac{\lambda(t)}{2}(\theta-\theta')^2}$ can be readily found and hence one immediately arrives at the following expression for $\bar{\rho}(t)$

\begin{equation}
\bar{\rho}(t)=e^{-it\mathcal{L}_{H}-\frac{\lambda(t)}{2}\mathcal{L}_{H}^{2}}[\rho(0)]
\end{equation}

\noindent where $\mathcal{L}_{H}\equiv[H,\cdot]$ is the commutator with the hamiltonian $H$. Note that this equation automatically satisfies condition i) above, i.e. $\bar{\rho}(t)$ is a density operator (it is selfadjoint, it has trace one and it is positive) provided the initial $\rho(0)$ is a density operator too. This is a first generalization of the results obtained in \cite{BonOliTomVit00a}. If condition ii) is imposed, then the only choice for $\sigma(t)$ will be $\sigma(t)=\gamma^{1/2}$, i.e. a constant. The master equation is the well-known phase-destroying master equation

\begin{equation}\label{PhasDestME}
\dot{\bar{\rho}}(t)=-i[H,\bar{\rho}(t)]-\frac{\gamma}{2}[H,[H,\bar{\rho}(t)]]
\end{equation}

Note that no approximation has been assumed to arrive at this equation, in contrast to \cite{BonOliTomVit00a,Mil91a}. The constant $\gamma$ is thus under this formalism  a measure of the deviation from strictly determinist time evolution. Apart from the evident freedom in the choice of adequately physically motivated choices of functions $b(\cdot,\cdot)$ and $\sigma(\cdot,\cdot)$ (which e.g. may drive us to time-dependent $\gamma(t)$), a higher generalization can be achieved if the previous scheme is slightly changed by substituting the whole phase of the exponential in (\ref{UnitOp}) for a $\theta$-dependent random variable $\chi_{t}(\theta)$. This allows us to write the random evolution operator instead as

\begin{equation}
U_{\text{rand}}(t)=\int_{-\pi}^{\pi}e^{-i\chi_{t}(\theta)}dE_{\theta}
\end{equation}

The former scheme is a particular case of this latter when $\chi_{t}(\theta)=\theta\chi_{t}$. The physical meaning behind this formalism is clear: the random time evolution does not affect on an equal footing to all energy levels of the system. 
Following then the same idea as above, we consider $\chi_{t}(t)$ as a solution to an Ito SDE, now $\theta$-dependent

\begin{equation}
d\chi_{t}(\theta)=b(t,\chi_{t}(\theta);\theta)dt+\sigma(t,\chi_{t}(\theta);\theta)d\mathcal{B}_{t}(\theta)
\end{equation}

\noindent i.e., now we have a SDE for each energy level. This must be supplemented with correlation properties among the brownian motions at different $\theta$'s. These will be expressed through the relation

\begin{equation}
d\mathcal{B}_{t}(\theta)d\mathcal{B}_{t}(\theta')=g(t;\theta,\theta')dt
\end{equation}

\noindent where $g(t;\theta,\theta')$ is basically for each $t$ a covariance function expressing the correlations among different $\mathcal{B}_{t}(\theta)$'s. From a physical standpoint it is clear that $g(t;\theta,\theta')=g(t;\theta',\theta)$ and since we are dealing with standard brownian motions $|g(t;\theta,\theta')|\leq 1$ for all $t,\theta,\theta'$.

It is difficult to proceed forward without making some assumptions. One may on one hand discuss on physical grounds particular choices for the different functions involved and find the solution or on the other hand look for general analytical properties which drives us to a sufficiently global situation. As before, as a first example, let us assign simple choices to $b(t,\chi_{t}(\theta);\theta)=h(\theta)$ and  $\sigma(t,\chi_{t}(\theta);\theta)=\sigma(t;\theta)$. The former corresponds to the fact that the unitary part of the dynamics is driven by a renormalized hamiltonian \cite{CalLeg83a}, as the following calculation shows; whereas the second states that the departure from deterministic evolution is caused by an uncorrelated noise with different intensity at different times for each $\theta$. Let us include some details. We have to evaluate the expression:

\begin{widetext}
\begin{equation}\label{RandEvolOpSevEn}
\bar{\rho}(t)=\int_{-\pi}^{\pi}\int_{-\pi}^{\pi}e^{-i(h(\theta)-h(\theta'))t}\mathbb{E}[e^{-i\left(\int_{0}^{t}\sigma(s;\theta)d\mathcal{B}_{s}(\theta)-\int_{0}^{t}\sigma(s;\theta')d\mathcal{B}_{s}(\theta')\right)}]dE_{\theta}\rho(0)dE_{\theta'}
\end{equation}
\end{widetext}

The expectation value is found using standard techniques. Define the stochastic process $X_{t}(\theta,\theta')$ by the SDE $dX_{t}(\theta,\theta')=\sigma(t;\theta)d\mathcal{B}_{t}(\theta)-\sigma(t;\theta')d\mathcal{B}_{t}(\theta')$; express $\exp(-iX_{t}(\theta,\theta'))$ as a power series; apply the Ito formula (cf. \cite{Oksendal98a}) to $f(X_{t})$ ($f(z)=z^n$); define $\beta_{n}(t;\theta,\theta')\equiv\mathbb{E}[X_{t}^{n}(\theta,\theta')]$ to find

\begin{subequations}
\begin{eqnarray}
\beta_{2n}(t;\theta,\theta')&=&\frac{(2n)!}{2^n n!}\lambda^{n}(t;\theta,\theta')\\
\beta_{2n+1}(t;\theta,\theta')&=&0
\end{eqnarray}
\end{subequations}

\noindent where $\lambda(t;\theta,\theta')\equiv\int_{0}^{t}\eta(s;\theta,\theta')ds$ and we have defined $\eta(t;\theta,\theta')=\sigma^{2}(t;\theta)g(t;\theta,\theta)+\sigma^{2}(t;\theta')g(t;\theta',\theta')-2\sigma(t;\theta)\sigma(t;\theta')g(t;\theta,\theta')$. Finally resum the series and then (\ref{RandEvolOpSevEn}) reduces to

\begin{equation}\label{RenNonMark}
\bar{\rho}(t)=\int_{-\pi}^{\pi}\int_{-\pi}^{\pi}e^{-it(h(\theta)-h(\theta'))}e^{-\frac{\lambda(t;\theta,\theta')}{2}}dE_{\theta}\rho(0)dE_{\theta'}
\end{equation}

We cannot proceed further without specifying $\sigma(t;\theta)$ and $g(t;\theta,\theta')$. To compare with the previous scheme let us define $\sigma(t;\theta)=\gamma^{1/2}\theta$ and assume a correlation function $g(t;\theta,\theta')=\exp(-\tau^2(\theta-\theta')^2)$. Here $\tau^{-1}$ roughly measures the distance of correlation in some arbitrary units. Then $\lambda(t;\theta,\theta')=\gamma t (\theta^{2}+\theta^{'2}-2\theta\theta'e^{-\tau^2(\theta-\theta')^{2}})$ and the master equation reads

\begin{eqnarray}
\dot{\bar{\rho}}(t)&=&-i[h(H),\bar{\rho}(t)]-\nonumber\\
&-&\frac{\gamma}{2}(H^{2}\bar{\rho}(t)+\bar{\rho}(t)H^{2}-2He^{-\tau^2\mathcal{L}_{H}^{2}}[\bar{\rho}(t)]H)
\end{eqnarray}

This is a CP markovian master equation\footnote{To convince oneself that this denotes a CP evolution, compare with eq. (4.1) in \cite{Lin76}.}  different from (\ref{PhasDestME}), though it reduces to it in the limit $\tau\to 0$, i.e. in the case of infinite-range correlations (all energy levels equally affected by the random evolution). In this way the formalism depicted here generalizes the approach in \cite{BonOliTomVit00a}, even when  restricting ourselves to the markovian regime. Indeed, whenever we may write $\lambda(t;\theta,\theta')=t\tilde{\lambda}(\theta,\theta')$ (which amounts to having $\sigma(t;\theta,\theta')$ and $g(t;\theta,\theta')$ time-independent) we arrive at a markovian evolution, since

\begin{eqnarray}
\bar{\rho}(t)&=&\int_{-\pi}^{\pi}\int_{-\pi}^{\pi}e^{t[-i(h(\theta)-h(\theta'))-\tilde{\lambda}(\theta,\theta')]}dE_{\theta}\rho(0)dE_{\theta'}\nonumber\\
&\equiv&e^{-it\mathcal{L}_{h(H)}+t\mathcal{D}_H}[\rho(0)]
\end{eqnarray}

\noindent where $\mathcal{L}_{h(H)}$ and $\mathcal{D}_H$ denote superoperators.\\
The previous example depicts a typical situation of nondissipative decoherence. In this language and restricting to the previous case (eq. (\ref{RenNonMark})), this is due to the fact that $\lambda(t,\theta,\theta)=0$, as an immediate calculation of $\text{Tr}(\bar{\rho}(t)H)$ shows. Working in the energy eigenstates basis, (\ref{RenNonMark}) leads to $\bar{\rho}(t;E,E')=\exp(-it(h(E)-h(E'))-\lambda(t;E,E')/2)\rho(0;E,E')$, hence 

\begin{eqnarray}
\text{Tr}(\rho(t)H)&=&\int_{\sigma(H)}Ee^{-\frac{\lambda(t;E,E)}{2}}\rho(0;E,E)d\mu(E)=\nonumber\\\label{MeanEn}
&=&\text{Tr}(\rho(0)H)
\end{eqnarray}

\noindent where $\sigma(H)$ denotes the spectrum of $H$ and $\mu(E)$ denotes the measure over this spectrum (a sum if it is discrete and an integral if it is continuous) and where $\lambda(t,\theta,\theta)=0$ has been assumed. From (\ref{MeanEn}) it is clear that if this condition is not satisfied, we have a dissipative decoherence process. This is a third generalization of previous works.\\
Some comments must be made about this formalism. Though it has been presented as an adequate tool to express decoherence in isolated systems, from a mathematical standpoint it can be viewed as an alternative description of environment-induced decoherence provided the stochastic parameter $\chi_{t}$ is properly connected to the surrounding reservoir properties and  the environment-system interaction characteristics. But in our opinion the main benefit from this language is to bring a bit closer the decoherence process and the different models of stochastic state reduction (see \cite{AdlBroBruHug01a} for a review of their common backbone). However this connection demands further mathematical work which is in progress. 

\begin{acknowledgments}
One of us (D.S.)  acknowledges the support of Madrid Education Council under grant BOCAM 20-08-1999. 
\end{acknowledgments}


\end{document}